\documentclass[12pt,preprint]{aastex}

\shorttitle{A young stellar population in the outer Third Galactic Quadrant}
\shortauthors{Carraro et al.}

\begin{document}


\title{Detection of a young stellar population in the background of 
open clusters in the  Third Galactic Quadrant}


\author{Giovanni Carraro\altaffilmark{1,2}}
\affil{Departamento de Astronom\'ia, Universidad de Chile, 
Casilla 36-D, Santiago, Chile}
\email{gcarraro@das.uchile.cl}

\author{Ruben A. V\'azquez}
\affil{               Facultad de Ciencias Astron\'omicas y Geof\'{\i}sicas de la
               UNLP, IALP-CONICET, Paseo del Bosque s/n 1900, La Plata, Argentina}
\email{rvazquez@fcaglp.fcaglp.unlp.edu.ar}

\author{Andr\'e Moitinho}
\affil{CAAUL, Observat\'orio Astron\'omico de Lisboa, Tapada da Ajuda,
               1349-018 Lisboa, Portugal}
\email{andre@oal.ul.pt}

\author{Gustavo Baume}
\affil{               Facultad de Ciencias Astron\'omicas y Geof\'{\i}sicas de la
               UNLP, IALP-CONICET, Paseo del Bosque s/n 1900, La Plata, Argentina}
\email{gbaume@fcaglp.fcaglp.unlp.edu.ar}

\altaffiltext{1}{Astronomy Department, Yale University, 
P.O. Box 208101, New Haven, CT 06520-8101 , USA}
\altaffiltext{2}{ANDES fellow, on leave from Dipartimento di Astronomia, Universit\`a di Padova,
Vicolo Osservatorio 2, I-35122, Padova, Italy.}


\begin{abstract}
We report the detection of a young stellar population ($\leq$100 Myrs) in the background
of 9 young open clusters belonging to a homogenoeous sample of 30 star clusters
in the Third Galactic Quadrant (at $217^o \leq l \leq 260^o$). 
Deep and accurate UBVRI photometry allows
us to measure model-independent age and distance for the clusters and the background population
with high confidence. This population is exactly
the same population (the Blue Plume) 
recently detected in 3 intermediate-age open clusters and suggested to be
a $\leq$ 1-2 Gyr old population belonging to the Canis Major (CMa) over-density (Bellazzini et al. 2004, 
Mart\'inez-Delgado et al. 2005). 
However, we find that the young population in those three and in six clusters of our sample
follows remarkably well the pattern of the Norma-Cygnus spiral arm as defined by CO clouds,
while in the other three program clusters it lies in the Perseus arm.
We finally provide one example (out of 21) of a cluster which does 
not show any background population,
demonstrating that this population is not ubiquitous toward CMa.
\end{abstract}


\section{Introduction}
The detection of the Canis Major (CMa) over-density by Martin et al. (2004)
produced a renaissance of interest for the Third Galactic Quadrant
($180^o \leq l \leq 270^0$) of the Milky Way (MW).
The lively debate in the last year (Momany et al.  2004, Bellazzini et al. 2004)
on this over-density clearly demands
a better picture of the Galaxy structure in this region. \\ 
For the last several years
our group has been  conducting a systematic 
homogeneous and accurate UBVRI photometric survey 
of Galactic open clusters in this
part of the Galaxy (see Moitinho (2001), Giorgi et al. (2005, and references therein), 
Carraro et al. (2005a, and references therein))
with the aim of understanding the detailed structure of the spiral
arm pattern in this quadrant. 
Young open clusters are recognized
as ideal spiral arm tracers (Becker \& Fenkart 1970, Feinstein 1994). Their young ages mean they are near the spiral arm
in which they formed and we can obtain
precise determinations of their reddening and distance, especially when deep
U band photometry - which is very effective in pinning down stars
with spectral type earlier than A0 - is available.\\
\noindent
Surprisingly, the shape and extent
of the Perseus and Cygnus-Norma  arms in the Third Quadrant are far from being
clear and settled. Russeil (2003) using star forming complexes
finds that both the Perseus and
Norma-Cygnus arms are not visible at all in the Third Quadrant, confirming
previous results by May et al. (1997) who mapped the region with CO clouds,
and at that time showed a lack of any
grand design spiral features in this Galaxy Location. Nevertheless, they could 
confirm previous suggestions about the shape and location of the Galactic warp
and show how bridges of material are present in a few anti-center
directions.
To date, no published study has probed the
spiral structure of the third Galactic Quadrant using young star clusters.
Modern surveys are therefore vital
to better trace the spiral pattern in this interesting,
but largely overlooked region of the Galaxy.\\

\noindent
In this Letter we report on a serendipitous result - namely the detection
of a young stellar population behind  a few young open clusters - we obtained
during the analysis of our large data-set of open clusters.

\section{Observational material} 
The CCD UBVRI photometry we are here using comes primarily from the Third Galactic Quadrant
survey ($217^o \leq l \leq 260^o$, $-5^o \leq b \leq +5^o$) described in full detail in
Moitinho (2001). 30 open clusters were observed with the CTIO 0.9m telescope
in two runs in 1994 and 1998. The data are homogeneous, with color errors of 0.1 mag in all combinations to
a typical limiting magnitude of V=21.
The analysis presented in Moitinho (2001) shows that the CTIO
photometry is accurate and consistent with other previously published
works.
For this reason we consider here  the 9 open clusters  NGC~2302, NGC~2383, NGC~2384,
NGC~2367, NGC~2362, NGC~2439, NGC~2533, NGC~2432 and Ruprecht~55 (see Table~1).
Eight of these clusters belong to the CTIO survey, whereas NGC~2362 was observed
with the 1.5 Danish telescope at La Silla in 2001. 
In all cases the field of view is 13$^{\arcmin} \times 13^{\arcmin}$ except for NGC2362,
which was surveyed with a mosaic of 5 $12^{\arcmin}.9 \times 13^{\arcmin}.3$ fields covering 540 arcmin
squared.
None of the
clusters have been investigated in much detail before, except NGC 2362
studied by Moitinho et al. (2001). They found that this cluster is nearby (1.5 kpc)
very young (5 million years) and exhibits a prominent pre- Main Sequence (MS) population.

\section{Photometric Diagrams} 
In Fig.~1 we show the Two Color Diagrams (TCDs) and Color-Magnitude Diagrams (CMDs)
of four open clusters : NGC~2302, NGC~2362,  NGC~2409 and NGC~2453.
We are going to use the first three
as templates in this work. The data we show in the plots have been selected
according to photometric errors, and in particular only the stars having 
$\sigma_{(U-B)}$, $\sigma_{(B-V)}$ and $\sigma_{V}$ simultaneously
 $\leq 0.09$ are shown. 
Although some data are available for these clusters, we remark this is the first time
deep and accurate CCD multiband photometry has been obtained and analyzed.
The full detailed analysis of these data will be presented in forthcoming
papers.\\

\noindent
Briefly, we follow the Zero Age Main Sequence (ZAMS) fitting method to
derive reddenings, distances and ages of stellar sequences in
photometric diagrams.  The TCDs allow us to infer the reddenings and the
spectral types of early type stars. The CMDs are then used to derive
each cluster's distance modulus and absolute distance. Finally, the
age is derived from the luminosity of the earliest spectral type star
still on the MS.
The fits were performed using the empirical solar metallicity Zero Age
Main Sequence (ZAMS) from Schmidt-Kaler (1982). The empirical ZAMS
does not suffer from the many uncertainties affecting theoretical
isochrones, and possible uncertainties due to chemical abundance are
negligible in early spectral type stars. The fits were done by eye,
yielding estimates of reddenings, distances and ages and of their
uncertainties.
This is quite a solid
method, traditionally used in studies of young star clusters. See
for reference Moffat (1971), Vogt \& Moffat (1972), Fitzgerald \&
Moffat (1980) or, more recently, Baume et al. (2004).

\noindent
A careful inspection of all the photometric diagrams reveals common 
signatures. In particular,  the presence of three distinct populations: 

\begin{itemize}

\item 
The {\it clusters population} revealed by the upper, bluer main sequences (MS).
These MSs are fitted in each case with the Schmidt-Kaler (1982) 
ZAMS properly shifted following a normal reddening
law, which we know to hold in this direction of the Galaxy (Moitinho 2001).
These fits are indicated in all diagrams as dotted lines.

\item A {\it fainter and more reddened young population} indicated by filled squares in
the NGC~2302 diagrams. 
We are going to refer to this population as the Blue Plume (BP). 
The ZAMS fits to the BP
are indicated in all the diagrams with solid lines.

\item The {\it Galactic disk field population}.

\end{itemize}

\noindent
The ZAMS fit allows us to obtain the basic parameters for the clusters and the BPs.
The results are summarized in Table~1, which lists for 
all the clusters their Galactic coordinates,
reddening estimates (column 4) for the whole line of sight from FIRB
maps of Schlegel et al. (1998), and their reddenings,
distances and ages derived from our fitting method (columns 5, 6 and
7).

\noindent
The BP is visible in  all the photometric diagrams of all the
clusters we mentioned in Section~2. In particular, in the TCDs of Fig~1 (upper 
panels) it is prominent in the case of NGC 2302 and a bit less detached from 
the cluster stars in NGC 2362 and NGC 2439. However all the 
corresponding CMDs (lower panels of Fig~1) show the BP presence beyond
any doubt.\\ 
As a comparison we also show in Fig~1 the TCD and CMD of the cluster 
NGC 2453 (right panel), where the BP is completely absent. 
It is remarkable that NGC~2453 ( $l  = 243^o.53$, $b = -0^o.93$) does
not show this feature.\\
The fit to the BPs (solid lines) provides us with  the
parameters for this population in different galactic directions (see columns 8, 9 and 10 in 
Table~1). 
In particular, we notice that the inferred helio-centric distances are always 
greater than 6 kpc, indicating that this population
is placed in the outskirts of 
the Galaxy. \\
On the other hand, the spectral types of the BP stars (in 
all the diagrams) go from B3-B5 to late A and F, suggesting that this population 
is actually young (ages lower than 100 Myr).

\section{Discussion and Conclusions}
In Fig.~2 and 3 we plot the location of the studied clusters (the 3 templates plus other 6 clusters,
see Table~1) with open squares and their BPs with solid triangles 
in the X-Y and X-Z planes.\\ 
We adopt filled circles to place  six more  clusters which are believed to be associated to CMa,
or which lie in the same region. They are one globular and two old open clusters (NGC~2298,
Tombaugh~2 and AM-2) taken from Frinchaboy et al. (2003) and three other old open clusters
(Berkeley 25, 73 and 75) taken from Carraro et al (2005a).\\

\noindent
To facilitate the interpretation we also include
the distribution of CO clouds from the recent study by May et al. (2005)
depicted as open circles. In this study, the authors provide
the result of a new large survey of CO clouds in the Third Galactic Quadrant
with high quality data taken at the Nanten telescope (Chile), showing that they trace very well
the expected position of the Norma-Cygnus spiral arm. \\
Many of these clouds 
harbor IRAS sources (Bronfman et al. 2005), suggesting that star formation is still on going 
at their location. It seems therefore that the existence of the Norma-Cygnus arm (often referred 
to as the outer arm)
in the Third Galactic Quadrant, previously not very clear, is a reality.
Moreover, the spiral arm extension is mostly detected low in the Galactic plane
at $+0.5\leq b \leq -6.50$ (May et al. 1997, 2005), being this a clear effect of the Galactic warp.\\

\noindent
The BP population of 6 out of 9 clusters closely follow distance, longitude and latitude of
the outer spiral arm in both the projections. The three clusters which deviate are NGC~2432, NGC~2533 and Ruprecht~55, which 
lie above the Galactic plane (see Fig.~3), are closer to the Sun, 
and follow the expected extension of the Perseus arm in
the Third Galactic Quadrant (Russeil 2003, May et al. 1997, 2005).

\noindent
It is remarkable that the BP is seen with the same age and shape in all the cluster fields 
irrespective of their position. It seems clear that the BP is a possible spiral arm indicator.\\

\noindent
Very interestingly, the BP we find appears also in the field of NGC~2477 (Momany et al. 2001),
Tombaugh~1 (Carraro \& Patat 1996) and Berkeley~33 (Carraro et al. 2005b), 
3 negative latitude open clusters (see Table~1),  and it is
currently suggested to be an intermediate-age ($\leq$ 1-2 Gyr) population associated
with the CMa over-density, or the last burst of star formation experienced
by this {\it galaxy}.\\
We estimated age and distance of the BP in these 3 clusters
(see Table~1),
and plot its position in Figs.~2 and 3 with starred symbols. The BP population in all these three
clusters was found to be young and to lie in the outer arm.\\

\noindent
Our findings support
the idea that the BP in the CMa (plotted with  a large filled square) 
direction is a young population mostly associated 
with the 
Norma-Cygnus arm, since it is  much younger than suggested before, more
distant and situated in  an area 
encompassing a significant sector (more than 40$^o$ in longitude) 
of the Third Quadrant, where the arm is expected to lie.\\

\noindent
Along this vein, an interesting consideration
comes from the inspection of the CMD of NGC~2168 ($l=186.59$, $b=+2.19$) 
by Kalirai et al. (2003, Fig.1). 
This cluster shows the BP, and the same MS as in NGC~2477, the F-XMM field
shown in Bellazzini et al. (2004) and the Martin\'ez-Delgado et al. (2005) deep CMD. 
Now, the CMa stream at the
position of NGC~2168 is far more distant and
at a different latitude.
We fitted an empirical ZAMS to the publicly available data, and 
find that the BP in NGC~2168, with a distance of 6.4$\pm$0.5 kpc, 
belongs to  the Norma-Cygnus arm.\\

\noindent
We also show
(Fig. 1) that NGC 2453 does not show the BP.  In fact, most of the clusters
in our sample from this region of the Galaxy do not show the BP.
This is remarkable, and means that the BP is not 
ubiquitous toward CMa area, as we would expect
for a young population associated to a galaxy, but it closely follows the structure
of the  Perseus and Norma-Cygnus spiral arms in the Third Galactic Quadrant. 

\noindent
In conclusion, our data support the idea that, whatever
the nature of CMa, the BP population is more consistent
with a young population in the Perseus and Norma-Cygnus spiral arms.

\noindent
As a final note, we surely agree that a significant stellar concentration is visible in the CMD
of NGC~2362 (see Fig~1) at V$\approx$19, remarkably similar to that seen in NGC~2477,
in the F-XMM field by Bellazzini et al. (2004) and in the deep CMD by Martin\'ez-Delgado et al.(2005).
This population lies between the cluster and its BP, so in the inter-arm zone, since
it is the same population ascribed to CMa, which lies at 8.5 kpc from the Sun.
By looking carefully at NGC~2362 CMD, we can notice 4 populations:

\begin{description}
\item $\bullet$ NGC 2362  at $-0.2 \leq (B-V) \leq 0.6$ ;
\item $\bullet$ the BP at $ 0.2 \leq (B-V) \leq 0.6$;
\item $\bullet$ the MW unevolved disk population at $ 0.4 \leq (B-V) \leq 1.1$;
\item $\bullet$ the disk giants at  $ 1.1 \leq (B-V) \leq 1.6$.
\end{description}

\noindent
Now, the extension of the CMa over-density is estimated to be around $40^o$ ($220^o \leq l \leq 260^o$)
and to have a line-of-sight depth of about 1 kpc (Mart\'inez-Delgado et al. 2005), which is
more or less the inter-arm separation between Perseus and Norma-Cygnus at
these longitudes (Russeil 2003).\\

\noindent
Intermediate-age and old stars are expected to populate the inter-arm region 
and both in our Galaxy and in external ones
bridges of matter connecting spiral arms are not rare. The Local arm, containing the Sun
is a nice example. May et al. (1997) actually show at $l \approx 240^o$ a clear
bridge of material extending for more than 6 kpc, exactly toward the CMa over-density direction.\\

\acknowledgments
We acknowledge very interesting discussions with Jorge May and Leonardo Bronfman
at Universidad de Chile, and thanks them very much for providing 
us with the results of their work before publication.
The work of G.C. is supported by {\it Fundaci\'on Andes}.
A.M. acknowledges support from FCT (Portugal) through grant
SFRH/BPD/19105/2004.
This work has been also developed in the framework of
the {\it Programa Cient\'ifico-Tecnol\'ogico Argentino-Italiano SECYT-MAE
C\'odigo: IT/PA03 - UIII/077 - per\'iodo 2004-2005}.
This study made use of Simbad and WEBDA databases.

\clearpage

\begin{figure}
\epsscale{1.1}
\plotone{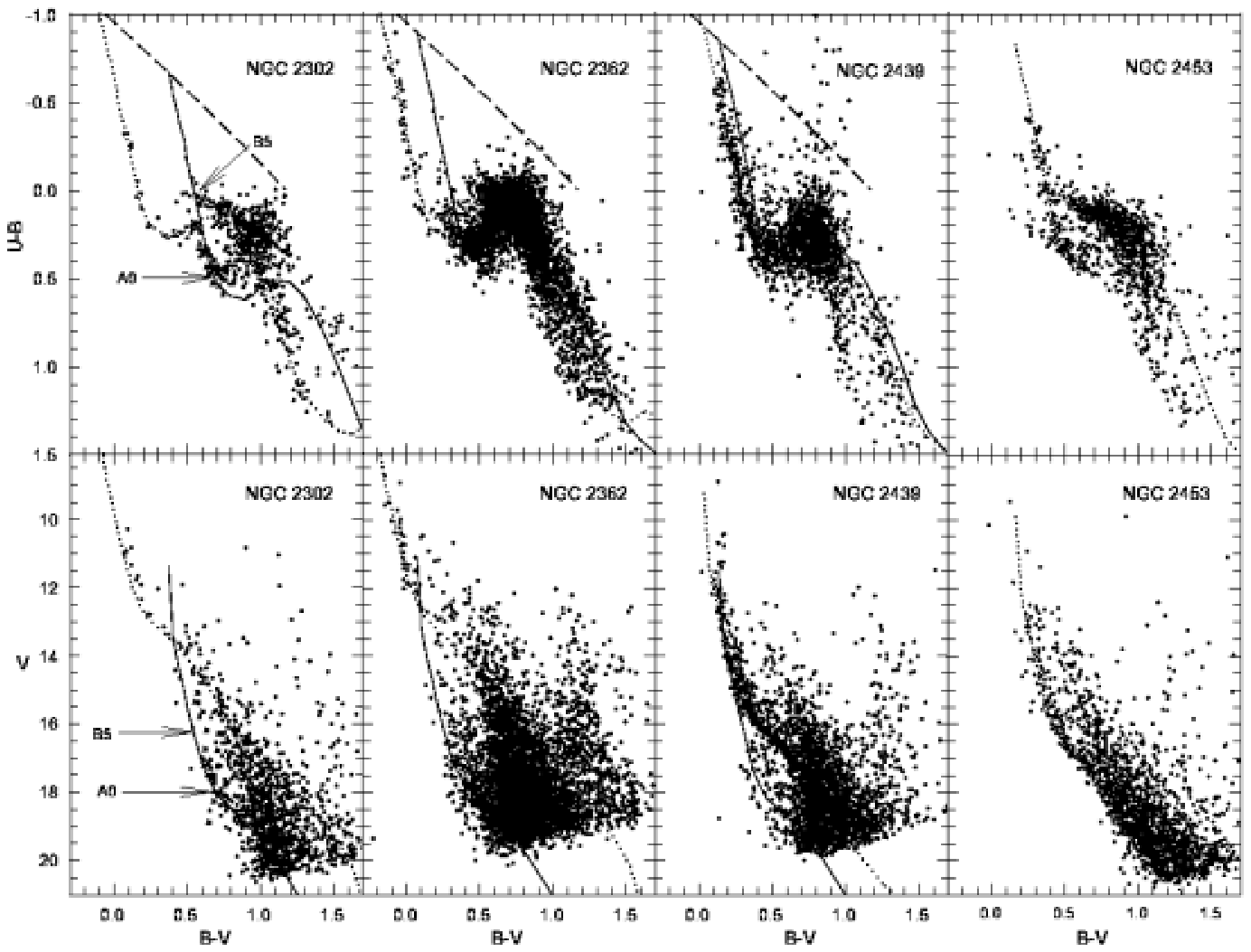}
\caption{TCDs and CMDs for the 3 template clusters NGC~2302, NGC~2362 and NGC~2439,  and for NGC~2453
All the continuous and dotted lines are Schmidt-Kaler empirical ZAMS. See text for further details.
The dashed lines in the TCDs indicate the reddening path.}
\end{figure}

\clearpage
\begin{figure}
\epsscale{1.0}
\plotone{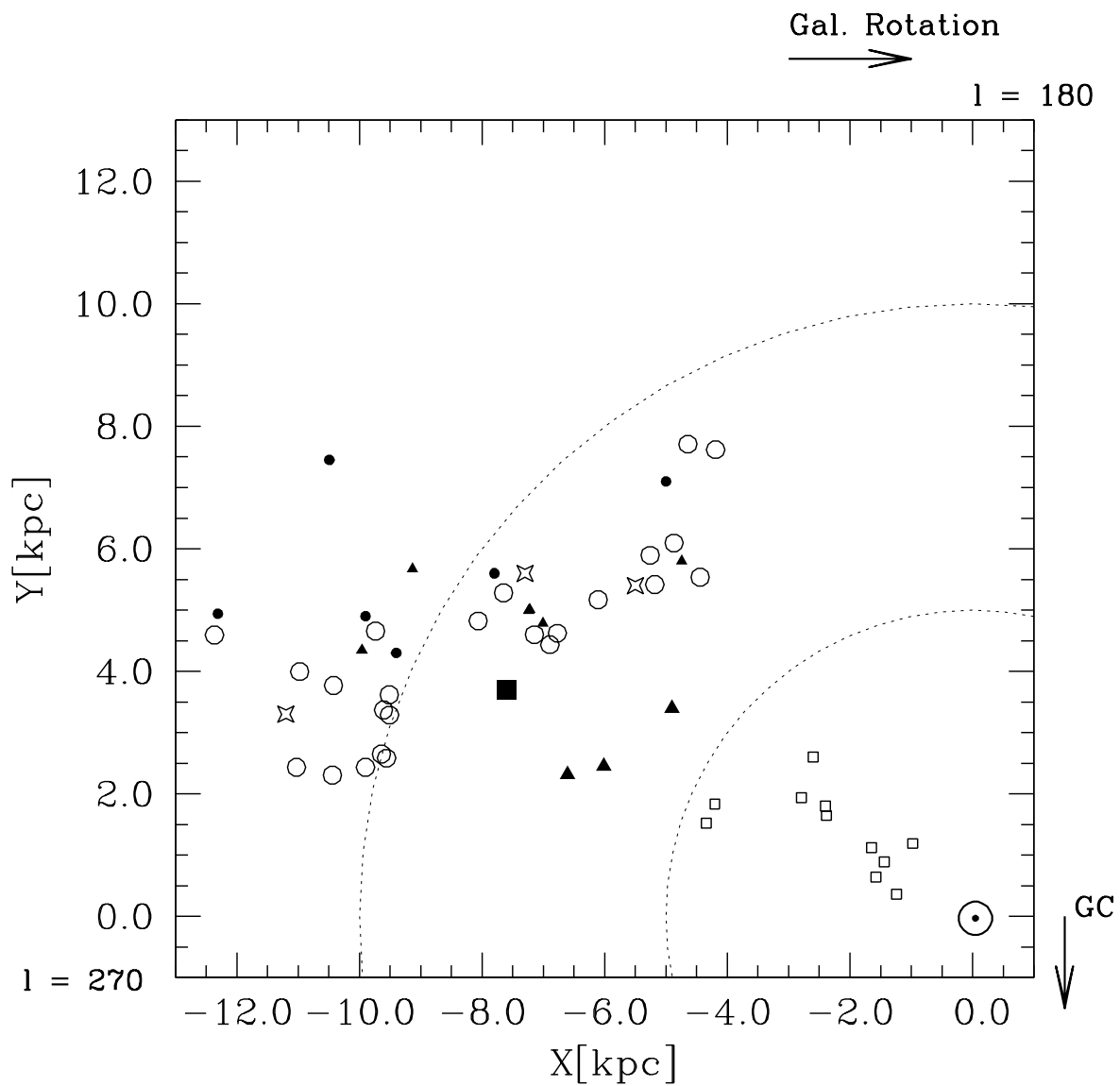}
\caption{A sketch of the Third Galactic Quadrant. The X-Y plane. Open squares: location of the open clusters
discussed in the paper. 
Filled triangles: location of the BP population in all the program clusters. Open circles:
location of the CO clouds by May et al. (2005). Starred symbols: location of the BP population
in NGC~2477, Tombaugh~1 and Berkeley~33. Large filled square: position of the CMa over-density.
Filled circles: location of  open and globular clusters suggested to be possibly associated with 
CM. At (0,0) the Sun position is indicated. 
The direction of the Galactic Center and rotation are indicated with solid arrows, whereas with the dotted 
simbols we show two constant heliocentric distance (5 and 10 kpc) circles.}
\end{figure}

\clearpage
\begin{figure}
\epsscale{1.0}
\plotone{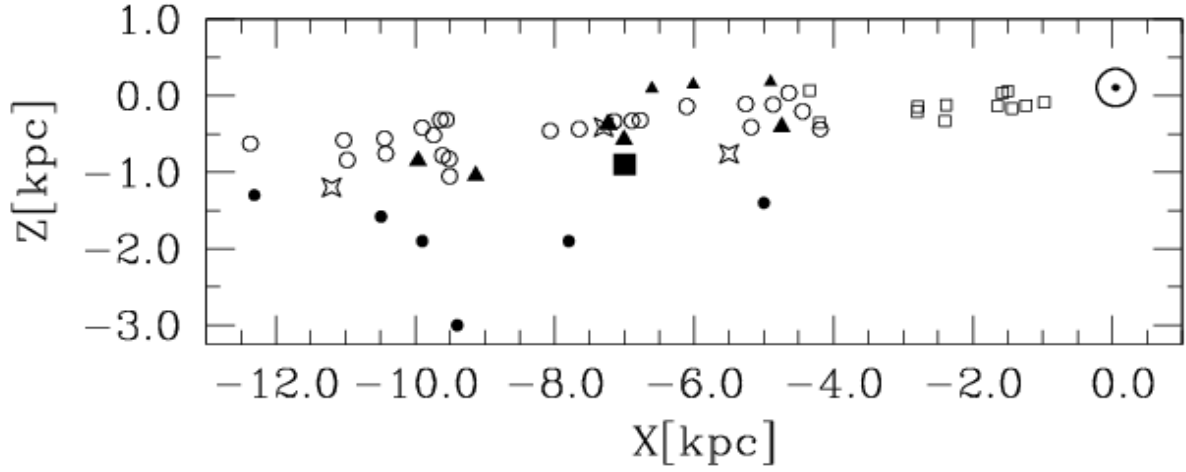}
\caption{A sketch of the Third Galactic Quadrant. The X-Z plane. Symbols as in Fig.~2.
The Sun is located at Z = +40 pc.}
\end{figure}

\clearpage
\begin{table*}
\caption{{}Basic properties of the clusters and the background population}
\fontsize{8} {10pt}\selectfont
\begin{tabular}{cccccccccc}
\hline
\multicolumn{1}{c}{$Name$} &
\multicolumn{1}{c}{$l$} &
\multicolumn{1}{c}{$b$} &
\multicolumn{1}{c}{$E(B-V)_{FIRB}$} &
\multicolumn{1}{c}{$E(B-V)$}  &
\multicolumn{1}{c}{${d_{\odot}}$} &
\multicolumn{1}{c}{$age$} &
\multicolumn{1}{c}{$E(B-V)_{BP}$}  &
\multicolumn{1}{c}{${d_{\odot,BP}}$} & 
\multicolumn{1}{c}{$age_{BP}$} \\
\hline
 & $^{o}$ & $^{o}$ & mag & mag & kpc & Myr & mag & kpc & Myr \\
\hline
NGC 2302     & 219.28 & -03.10 & 0.84 & 0.23 & 1.5 &   12 & 0.70$\pm$0.05 &  7.5$\pm$0.5 & $\leq$ 100\\
NGC 2383     & 235.27 & -02.43 & 0.73 & 0.30 & 3.4 &  120 & 0.56$\pm$0.05 &  8.8$\pm$0.5 & $\leq$ 100\\
NGC 2384     & 235.39 & -02.42 & 0.72 & 0.29 & 2.9 &   12 & 0.56$\pm$0.05 &  8.8$\pm$0.5 & $\leq$ 100\\
NGC 2367     & 235.64 & -03.85 & 1.07 & 0.05 & 1.4 &    5 & 0.62$\pm$0.05 &  8.5$\pm$0.5 & $\leq$ 100\\
NGC 2362     & 238.18 & -05.55 & 0.53 & 0.13 & 1.4 &    5 & 0.40$\pm$0.15 & 10.8$\pm$1.2 & $\leq$ 100\\
NGC 2439     & 246.41 & -04.43 & 0.58 & 0.37 & 1.3 &   10 & 0.47$\pm$0.10 & 10.9$\pm$1.1 & $\leq$ 100\\
NGC 2533     & 247.80 & +01.29 & 0.67 & 0.14 & 1.7 &  700 & 0.48$\pm$0.05 &  6.5$\pm$0.3 & $\leq$ 100\\
NGC 2432     & 235.48 & +01.78 & 0.76 & 0.23 & 1.9 &  500 & 0.48$\pm$0.10 &  6.0$\pm$0.5 & $\leq$ 100\\
Ruprecht 55  & 250.68 & +00.76 & 0.80 & 0.45 & 4.6 &   10 & 0.50$\pm$0.05 &  7.0$\pm$0.5 & $\leq$ 100\\
\hline
\hline
NGC 2477     & 253.56 & -05.84 & 0.65 & 0.24 & 1.3 &  600 & 0.56$\pm$0.10 & 11.7$\pm$1.0 & $\leq$ 100 \\
Berkeley 33  & 225.40 & -03.12 & 0.80 & 0.47 & 4.0 &  800 & 0.61$\pm$0.10 &  7.7$\pm$0.5 & $\leq$ 100 \\
Tombaugh 1   & 232.33 & -06.31 & 0.54 & 0.40 & 3.0 & 1000 & 0.52$\pm$0.10 &  7.7$\pm$0.5 & $\leq$ 100 \\    
\hline
\end{tabular}
\end{table*}

\end{document}